\documentclass[trackchanges]{aastex701}
\usepackage{booktabs,longtable,threeparttablex}   
\usepackage{placeins}

\begin{document}

\title{Photometric and Astrometric Information for Sources around HD~163296 Revealed by JWST/NIRCam Coronagraphy}

\correspondingauthor{Luca Ricci}
\email{luca.ricci@csun.edu}

\author[0000-0002-6879-3030]{Taichi Uyama}
    \affiliation{Astrobiology Center, 2-21-1 Osawa, Mitaka, Tokyo 181-8588, Japan}
    \affiliation{National Astronomical Observatory of Japan, 2-21-1 Osawa, Mitaka, Tokyo 181-8588, Japan}
    \email{taichi.uyama.astro@gmail.com}
\author[0000-0001-8123-2943]{Luca Ricci}
    \affiliation{Department of Physics and Astronomy, California State University Northridge, 18111 Nordhoff Street, Northridge, CA 91330, USA}
    \email{luca.ricci@csun.edu}
\author[0000-0001-7591-2731]{Marie Ygouf}
    \affiliation{Jet Propulsion Laboratory, California Institute of Technology, 4800 Oak Grove Dr., Pasadena, CA, 91109, USA}
    \email{marie.ygouf@jpl.nasa.gov}
\author[0000-0002-9573-3199]{Massimo Robberto}
    \affiliation{Space Telescope Science Institute, 3700 San Martin Drive, Baltimore, MD 21218, USA}
    \affiliation{Department of Physics \& Astronomy, Johns Hopkins University, 3400 N. Charles Street, Baltimore, MD 21218, USA}
    \email{robberto@stsci.edu}

\begin{abstract}

Background stars observed through a circumstellar disk provide valuable benchmarks for investigating the disk's extinction properties. The HD~163296 system is an excellent case study due to its large disk, the clearly visible extinction effects in JWST/NIRCam data, and the presence of numerous background sources within or around its disk. We present the measured contrasts and astrometry of sources surrounding HD~163296 from Cycle~1 JWST/NIRCam coronagraphic observations, which will serve as a useful reference for future studies of the disk's extinction characteristics.

\end{abstract}

\section{Introduction}

Direct imaging has the potential to detect planets in protoplanetary disks and test the predictions of planet-disk interaction models \citep[e.g.,][]{Bae2023}. However, the current direct detection yield has been small \citep[e.g., PDS~70bc, WISPIT2-b;][]{Keppler2018,Haffert2019,vanCapelleveen2025}.
Different scenarios can explain the non-detections of the observational surveys \citep[e.g.,][]{Asensio-Torres2021,Cugno2023,Wallack2024}. One explanation is related to the large uncertainty on the intrinsic luminosity of protoplanets, with models predicting very different values \citep[e.g., different initial planet entropy - hot/warm/cold-start;][]{Spiegel2012}. Another possibility is significant extinction of the protoplanet light due to dust in the protoplanetary disk \citep[e.g.,][]{Sanchis2020,Alarcon2024}. However, these hypotheses are difficult to test due to insufficient observational constraints. 

Recently, \cite{Cugno2025} used spectrophotometry of a background star behind the AS~209 protoplanetary disk to measure the extinction law of dust in this system. This unique method is applicable to any background star that might suffer from disk extinction.
Applying a similar perspective, \cite{Uyama2025} identified regions of significant extinction caused by the HD~163296 disk at $\lambda\approx2\mu{\rm m}$, extending out to $\sim3"$ from the central star.
Given the outer radius of $\sim5"$ for the CO gas in this disk \citep{Isella2018} and numerous background sources around HD~163296, which could suffer from extinction by the HD~163296 disk, this is one of the best systems to directly measure the extinction by disk.
Here we report photometric and astrometric information of background sources detected by the JWST/NIRCam observations presented in \cite{Uyama2025}, which can be used for future observational studies to investigate the extinction effect in detail.

\section{Data}

We used the JWST/NIRCam F200W+MASK210R and F410+MASK430R data of HD~163296 \cite[GO 2540, PI: Luca Ricci;][]{Uyama2025}. 
For data reduction, we utilized roll-subtraction angular differential imaging technique \citep[ADI;][]{Marois2006} with the spaceKLIP post-processing pipeline \citep{Kammerer2022,Carter2023} and adopted the same configuration in spaceKLIP as \cite{Uyama2025}. Note that the output FoV was cropped into $\sim10"\times10"$, where HD~163296 is centered in the cropped FoV. 

After ADI post-processing, we used the {\tt analysistools.extract\_companions} function in spaceKLIP, which conducts forward modeling and PSF fitting \citep[][]{Pueyo2016}, to extract contrast and relative astrometry of the sources from the central star.
We note that the diffraction patterns of NIRCam, 'negative side-lobes' that are self-subtraction artifacts due to ADI reduction, and the fact that these features are complicatedly mixed due to the dense population, limit the accuracy of PSF fitting of most sources (see Figure~\ref{fig: F200W}). 
Therefore we conducted a single series of PSF fitting from bright sources to faint sources, where forward-modeled PSFs are removed before PSF fitting of a next source. In this process we also fixed a small box-size for PSF fitting (13x13-pix for F200W and 11x11-pix for F410M) to avoid the effects of nearby sources. We prioritized avoiding fatal errors when running {\tt analysistools.extract\_companions} in the whole analysis series. The information provided here can guide future studies targeting specific background stars by using the astrometric relative measurements, as well as our photometric measurements as initial estimates that can be improved on a source-by-source case.

\section{Results}

\begin{figure}[h]
    \centering
    \includegraphics[width=0.93\linewidth]{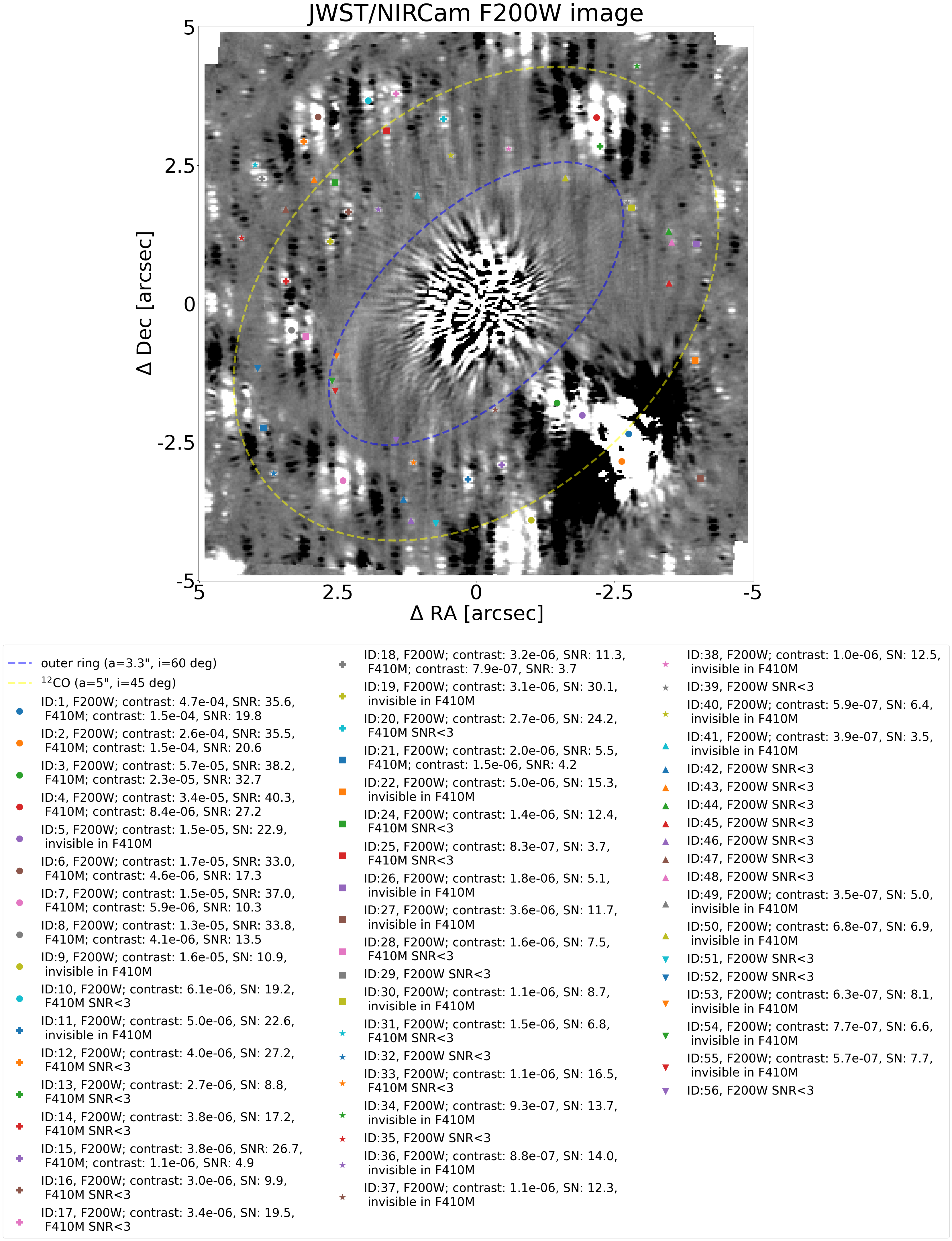}
    \caption{Identified sources around HD~163296 together with contrast and SNR information. Also highlighted are the locations of the outer ring feature resolved by the NIRCam observations \citep[blue line;][]{Uyama2025} and the outer edge of the $^{12}{\rm CO}$ gas \citep[yellow line;][]{Isella2018}.
    The extracted contrast and relative astrometry is available in Table~\ref{tab:bg_sources_grouped} (this table is available in the arXiv version only), also in machine-readable form in the original article. 
    }
    \label{fig: F200W}
\end{figure}

Figure~\ref{fig: F200W} shows the sources we identified in the JWST data 
and Table~\ref{tab:bg_sources_grouped} summarizes the contrast and astrometry. We confirmed that the extracted relative astrometries are consistent with visual locations of the sources. As mentioned above, we fixed the box size for fitting and SNRs can be improved by changing such configurations for each individual source.
We did not convert contrast values into apparent magnitudes using catalog values for the central star photometric information. There is no photometric reference star in the NIRCam FoV except for HD~163296 that is variable as it is a Herbig Ae/Be star, and presenting contrast provides more reliable information without uncertain systematics due to HD~163296's variability.
Note that we have missed extracting $\sim10$ sources, which we visually recognized in the post-processed image, because of the effect of nearby sources (i.e. diffraction patterns and ADI side-lobes) or faintness of the sources (SNR$<3$ after PSF fitting).
Also, sources close to the edge of the cropped FoV could not be fitted. 

As mentioned in \cite{Mesa2019} and \cite{Uyama2025}, all sources are most likely background objects and several tens of the sources are located behind the disk. Given the proper motion of HD~163296 \citep[$\mu_{\rm RA}=-7.59\pm0.04\ {\rm mas}, \mu_{\rm Dec}=-39.46\pm0.03\ {\rm mas}$;][]{GaiaDR3}, background sources currently located south of HD~163296 will suffer from disk extinction in the future.

\FloatBarrier
\begin{ThreePartTable}
\begin{TableNotes}
\item[a] This table is available in the arXiv version only. The machine-readable table is available in the original article.
\item[b] The sources in the F200W image where we had fatal errors in {\tt analysistools.extract\_companions} are not presented. '-' in the F410M SNR column indicates that the source is not recognized by eye. '$<3$' indicates that it is visible but PSF fitting resulted in a low SNR.
\end{TableNotes}

\begin{longtable}{rcccccccc}
\caption{Summary of Extracted Sources$^a$} \label{tab:bg_sources_grouped} \\
\toprule
 & \multicolumn{4}{c}{F200W} & \multicolumn{4}{c}{F410M} \\
ID & $\Delta$RA & $\Delta$Dec & contrast & SNR & $\Delta$RA & $\Delta$Dec & contrast & SNR \\
\midrule
\endfirsthead
\caption[]{Summary of Extracted Sources$^a$} \\
\toprule
 & \multicolumn{4}{c}{F200W} & \multicolumn{4}{c}{F410M} \\
ID & $\Delta$RA & $\Delta$Dec & contrast & SNR & $\Delta$RA & $\Delta$Dec & contrast & SNR$^b$ \\
\midrule
\endhead
\midrule
\multicolumn{9}{r}{Continued on next page} \\
\midrule
\endfoot
\bottomrule
\endlastfoot
1 & $-2\farcs754 \pm 0\farcs001$ & $-2\farcs3535 \pm 0\farcs0005$ & 4.74e-04 & 35.6 & $-2\farcs737 \pm 0\farcs001$ & $-2\farcs360 \pm 0\farcs001$ & 1.52e-04 & 19.8 \\
2 & $-2\farcs630 \pm 0\farcs001$ & $-2\farcs8473 \pm 0\farcs0004$ & 2.63e-04 & 35.5 & $-2\farcs610 \pm 0\farcs001$ & $-2\farcs856 \pm 0\farcs001$ & 1.55e-04 & 20.6 \\
3 & $-1\farcs460 \pm 0\farcs001$ & $-1\farcs7919 \pm 0\farcs0004$ & 5.66e-05 & 38.2 & $-1\farcs454 \pm 0\farcs001$ & $-1\farcs797 \pm 0\farcs001$ & 2.31e-05 & 32.7 \\
4 & $-2\farcs175 \pm 0\farcs002$ & $3\farcs364 \pm 0\farcs001$ & 3.40e-05 & 40.3 & $-2\farcs209 \pm 0\farcs015$ & $3\farcs370 \pm 0\farcs010$ & 8.40e-06 & 27.2 \\
5 & $-1\farcs918 \pm 0\farcs001$ & $-2\farcs016 \pm 0\farcs001$ & 1.48e-05 & 22.9 & -- & -- & -- & -- \\
6 & $2\farcs855 \pm 0\farcs002$ & $3\farcs375 \pm 0\farcs002$ & 1.67e-05 & 33.0 & $2\farcs824 \pm 0\farcs009$ & $3\farcs381 \pm 0\farcs006$ & 4.60e-06 & 17.3 \\
7 & $2\farcs406 \pm 0\farcs003$ & $-3\farcs197 \pm 0\farcs002$ & 1.48e-05 & 37.0 & $2\farcs387 \pm 0\farcs050$ & $-3\farcs203 \pm 0\farcs030$ & 5.89e-06 & 10.3 \\
8 & $3\farcs338 \pm 0\farcs003$ & $-0\farcs480 \pm 0\farcs002$ & 1.29e-05 & 33.8 & $3\farcs322 \pm 0\farcs021$ & $-0\farcs487 \pm 0\farcs012$ & 4.09e-06 & 13.5 \\
9 & $-0\farcs994 \pm 0\farcs006$ & $-3\farcs911 \pm 0\farcs004$ & 1.61e-05 & 10.9 & -- & -- & -- & -- \\
10 & $1\farcs950 \pm 0\farcs005$ & $3\farcs671 \pm 0\farcs003$ & 6.07e-06 & 19.2 & -- & -- & -- & $<3$ \\
11 & $0\farcs147 \pm 0\farcs003$ & $-3\farcs169 \pm 0\farcs002$ & 4.95e-06 & 22.6 & -- & -- & -- & -- \\
12 & $3\farcs118 \pm 0\farcs006$ & $2\farcs931 \pm 0\farcs004$ & 3.99e-06 & 27.2 & -- & -- & -- & $<3$ \\
13 & $-2\farcs239 \pm 0\farcs030$ & $2\farcs847 \pm 0\farcs020$ & 2.71e-06 & 8.8 & -- & -- & -- & $<3$ \\
14 & $3\farcs437 \pm 0\farcs013$ & $0\farcs407 \pm 0\farcs008$ & 3.80e-06 & 17.2 & -- & -- & -- & $<3$ \\
15 & $-0\farcs463 \pm 0\farcs003$ & $-2\farcs910 \pm 0\farcs001$ & 3.79e-06 & 26.7 & $-0\farcs446 \pm 0\farcs029$ & $-2\farcs905 \pm 0\farcs028$ & 1.08e-06 & 4.9 \\
16 & $2\farcs308 \pm 0\farcs002$ & $1\farcs659 \pm 0\farcs001$ & 2.99e-06 & 9.9 & -- & -- & -- & $<3$ \\
17 & $1\farcs446 \pm 0\farcs014$ & $3\farcs791 \pm 0\farcs008$ & 3.44e-06 & 19.5 & -- & -- & -- & $<3$ \\
18 & $3\farcs872 \pm 0\farcs014$ & $2\farcs258 \pm 0\farcs008$ & 3.25e-06 & 11.3 & $3\farcs925 \pm 0\farcs064$ & $2\farcs300 \pm 0\farcs074$ & 7.95e-07 & 3.7 \\
19 & $2\farcs637 \pm 0\farcs002$ & $1\farcs125 \pm 0\farcs001$ & 3.06e-06 & 30.1 & -- & -- & -- & -- \\
20 & $0\farcs587 \pm 0\farcs014$ & $3\farcs335 \pm 0\farcs010$ & 2.70e-06 & 24.2 & -- & -- & -- & $<3$ \\
21 & $3\farcs847 \pm 0\farcs016$ & $-2\farcs246 \pm 0\farcs013$ & 1.99e-06 & 5.5 & $3\farcs792 \pm 0\farcs024$ & $-2\farcs240 \pm 0\farcs030$ & 1.50e-06 & 4.2 \\
22 & $-3\farcs954 \pm 0\farcs015$ & $-1\farcs027 \pm 0\farcs010$ & 4.98e-06 & 15.3 & -- & -- & -- & -- \\
23 & $-3\farcs412 \pm 0\farcs010$ & $3\farcs149 \pm 0\farcs007$ & 1.82e-06 & 14.7 & -- & -- & -- & -- \\
24 & $2\farcs551 \pm 0\farcs022$ & $2\farcs188 \pm 0\farcs020$ & 1.45e-06 & 12.4 & -- & -- & -- & $<3$ \\
25 & $1\farcs620 \pm 0\farcs046$ & $3\farcs122 \pm 0\farcs044$ & 8.32e-07 & 3.7 & -- & -- & -- & $<3$ \\
26 & $-3\farcs978 \pm 0\farcs034$ & $1\farcs081 \pm 0\farcs021$ & 1.75e-06 & 5.1 & -- & -- & -- & -- \\
27 & $-4\farcs051 \pm 0\farcs005$ & $-3\farcs153 \pm 0\farcs003$ & 3.55e-06 & 11.7 & -- & -- & -- & -- \\
28 & $3\farcs078 \pm 0\farcs008$ & $-0\farcs592 \pm 0\farcs005$ & 1.60e-06 & 7.5 & -- & -- & -- & $<3$ \\
29 & $3\farcs753 \pm 0\farcs051$ & $-0\farcs070 \pm 0\farcs059$ & -- & $<3$ & -- & -- & -- & -- \\
30 & $-2\farcs805 \pm 0\farcs029$ & $1\farcs731 \pm 0\farcs020$ & 1.10e-06 & 8.7 & -- & -- & -- & -- \\
31 & $3\farcs997 \pm 0\farcs014$ & $2\farcs507 \pm 0\farcs009$ & 1.53e-06 & 6.8 & -- & -- & -- & $<3$ \\
32 & $3\farcs658 \pm 0\farcs042$ & $-3\farcs065 \pm 0\farcs035$ & -- & $<3$ & -- & -- & -- & $<3$ \\
33 & $1\farcs134 \pm 0\farcs009$ & $-2\farcs870 \pm 0\farcs006$ & 1.06e-06 & 16.5 & -- & -- & -- & $<3$ \\
34 & $-2\farcs907 \pm 0\farcs008$ & $4\farcs293 \pm 0\farcs005$ & 9.29e-07 & 13.7 & -- & -- & -- & -- \\
35 & $4\farcs238 \pm 0\farcs060$ & $1\farcs188 \pm 0\farcs057$ & -- & $<3$ & -- & -- & -- & -- \\
36 & $1\farcs770 \pm 0\farcs010$ & $1\farcs702 \pm 0\farcs006$ & 8.77e-07 & 14.0 & -- & -- & -- & -- \\
37 & $-0\farcs341 \pm 0\farcs005$ & $-1\farcs919 \pm 0\farcs003$ & 1.08e-06 & 12.3 & -- & -- & -- & -- \\
38 & $-0\farcs587 \pm 0\farcs005$ & $2\farcs798 \pm 0\farcs003$ & 1.01e-06 & 12.5 & -- & -- & -- & -- \\
39 & $-2\farcs729 \pm 0\farcs055$ & $1\farcs826 \pm 0\farcs061$ & -- & $<3$ & -- & -- & -- & -- \\
40 & $0\farcs461 \pm 0\farcs010$ & $2\farcs679 \pm 0\farcs007$ & 5.95e-07 & 6.4 & -- & -- & -- & -- \\
41 & $1\farcs067 \pm 0\farcs023$ & $1\farcs962 \pm 0\farcs017$ & 3.90e-07 & 3.5 & -- & -- & -- & -- \\
42 & $1\farcs314 \pm 0\farcs058$ & $-3\farcs534 \pm 0\farcs058$ & -- & $<3$ & -- & -- & -- & -- \\
43 & $2\farcs930 \pm 0\farcs064$ & $2\farcs245 \pm 0\farcs057$ & -- & $<3$ & -- & -- & -- & -- \\
44 & $-3\farcs479 \pm 0\farcs061$ & $1\farcs309 \pm 0\farcs062$ & -- & $<3$ & -- & -- & -- & -- \\
45 & $-3\farcs485 \pm 0\farcs065$ & $0\farcs372 \pm 0\farcs063$ & -- & $<3$ & -- & -- & -- & -- \\
46 & $1\farcs179 \pm 0\farcs060$ & $-3\farcs913 \pm 0\farcs056$ & -- & $<3$ & -- & -- & -- & -- \\
47 & $3\farcs441 \pm 0\farcs067$ & $1\farcs707 \pm 0\farcs063$ & -- & $<3$ & -- & -- & -- & -- \\
48 & $-3\farcs531 \pm 0\farcs062$ & $1\farcs110 \pm 0\farcs057$ & -- & $<3$ & -- & -- & -- & -- \\
49 & $1\farcs555 \pm 0\farcs018$ & $2\farcs514 \pm 0\farcs016$ & 3.53e-07 & 5.0 & -- & -- & -- & -- \\
50 & $-1\farcs613 \pm 0\farcs008$ & $2\farcs269 \pm 0\farcs005$ & 6.82e-07 & 6.9 & -- & -- & -- & -- \\
51 & $0\farcs730 \pm 0\farcs063$ & $-3\farcs972 \pm 0\farcs060$ & -- & $<3$ & -- & -- & -- & -- \\
52 & $3\farcs947 \pm 0\farcs061$ & $-1\farcs171 \pm 0\farcs062$ & -- & $<3$ & -- & -- & -- & -- \\
53 & $2\farcs527 \pm 0\farcs014$ & $-0\farcs947 \pm 0\farcs008$ & 6.28e-07 & 8.1 & -- & -- & -- & -- \\
54 & $2\farcs604 \pm 0\farcs010$ & $-1\farcs396 \pm 0\farcs011$ & 7.74e-07 & 6.6 & -- & -- & -- & -- \\
55 & $2\farcs546 \pm 0\farcs016$ & $-1\farcs574 \pm 0\farcs010$ & 5.67e-07 & 7.7 & -- & -- & -- & -- \\
56 & $1\farcs446 \pm 0\farcs060$ & $-2\farcs463 \pm 0\farcs061$ & -- & $<3$ & -- & -- & -- & -- \\

\insertTableNotes
\end{longtable}
\end{ThreePartTable}

\FloatBarrier

\begin{acknowledgments}

Support for program GO~2540 was provided by NASA through a grant from the Space Telescope Science Institute, which is operated by the Association of Universities for Research in Astronomy, Inc., under NASA contract NAS 5-03127. Part of this work was carried out at the Jet Propulsion Laboratory, California Institute of Technology, under a contract with the National Aeronautics and Space Administration (80NM0018D0004).
This work is based on observations made with the NASA/
ESA/CSA James Webb Space Telescope. 
The JWST data presented in this article were obtained from the Mikulski Archive for Space Telescopes (MAST) at the Space Telescope Science Institute, which is operated by the Association of Universities for Research in Astronomy, Inc., under NASA contract NAS 5-03127 for JWST. The specific observations analyzed can be accessed via \dataset[doi: 10.17909/j87q-jf82]{https://archive.stsci.edu/doi/resolve/resolve.html?doi=10.17909/j87q-jf82}.

\end{acknowledgments}

\bibliography{reference}{}
\bibliographystyle{aasjournalv7}



\end{document}